\documentclass{article}
\usepackage{amsmath}
\begin{document}
\def\half{{\textstyle\frac{1}{2}}}
%

\title{FROM QUANTUM DEFORMATIONS OF RELATIVISTIC SYMMETRIES 
TO MODIFIED KINEMATICS AND DYNAMICS  }

\author{JERZY LUKIERSKI\\
Institute for Theoretical Physics\\
University of Wroc{\l}aw, pl. Maxa Borna 8,
\\
50-204 Wroc{\l}aw
Poland \\
lukier@ift.uni.wroc.pl}
\date{}
\maketitle

\begin{abstract}
Starting from noncommutative generalization of Minkowski space we consider quantum deformed relativistic symmetries which lead to the modification of kinematics of special relativity. The noncommutative field theory framework described by means of the star product formalism is briefly described. We briefly present the quantum modifications of Einstein gravity
\end{abstract}


\section{Introduction}	
The classical picture of space-time locally described by flat
Minkowski space $R^{3,1}$ with numerical coordinates 
$x_\mu =(x_0, \vec{x})$ is expected to break down at extremely short distances of the order of Planck length
$\lambda_{p} = \left( \frac{G\hbar}{c^3}\right)^{\half}
\simeq 10^{-33}$cm. The mechanism implying this modification are the quantum gravity effects causing the gravitational collapse in the measuring process of the distances below the Planck scale \cite{lukchin1,lukchin2}. It follows that operationally the space-time in the presence of quantized gravitational field becomes noncommutative. The quantum structure of space-time in the form of nonvariantly commutates coordinates of space-time events, as derived from Heinsenberg`s uncertainty principle and Einstein gravity equations, has been firstly obtained in \cite{lukchin2}.

At present there is usually considered the following class of noncommutative space-time coordinates $\widehat{x}_{\mu}$
\begin{equation}\label{luchi1}
[\widehat{x}_{\mu}, \widehat{x}_{\nu}]
= i \lambda^2_{p} \, \theta_{\mu\nu} 
+ i \lambda_{p} \, \theta_{\mu\nu}^{\quad \rho}
\widehat{x}_\rho
\end{equation}
where  $\theta_{\mu\nu}^{\quad \rho}$, $\theta_{\mu\nu}^{\quad \rho}$ are constant dimensionless tensors. The first term on rhs of Eq.~(\ref{luchi1}) describes the DFR canonical noncommutativity, and the second one is the Lie-algebraic deformation, with the most studied special case
described be so-called $\kappa$-deformation  \cite{lukchin3,lukchin4,lukchin5} ($\kappa\equiv (a \lambda_{p})^{-1})$
\begin{equation}\label{luchi2}
[\widehat{x}_{i}, \widehat{x}_{0}]
= i \lambda_{p} \, \widehat{x}_\nu
\qquad
[\widehat{x}_{i}, \widehat{x}_{j}]
= 0\, .
\end{equation}
It should be added that the noncommutativity of space-time coordinates has been as well derived in the context of quantized open string theory  \cite{lukchin6}.

In special relativity framework the space-time coordinates are described by the translation sector of the Poincar\'{e} group. If we introduce noncommutative space-time coordinates, such approach implies that Poincar\'{e} symmetries should be modified into quantum Poincar\'{e} symmetries, with noncommutative group parameters. The deformations  
Eqs.~(\ref{luchi1}-\ref{luchi2})
introduce the fundamental mass $\kappa$ with its universe as deformation parameter. It is physically very appealing that quantum relativistic symmetries may introduce besides $c$ (light velocity) and $\hbar$ (Planck constant) the third fundamental constant interpreting the Newton constant $G$, or equivalently the Planck mass $m_p = \sqrt{\frac{\hbar c}{G}} \simeq 2.2\cdot
10^{-5}g \simeq 1.2 \cdot 10^{19} \frac{GeV}{c^2}$.
Quantum relativistic symmetries, with noncommutative space-time and symmetry generators satisfying deformed Lie-algebraic relations, are described by dually related pair of Hopf algebras \cite{lukchin7}. Analogously as in undeformed case, when we pass by exponential map from Lie-algebraic classical symmetry generators to the elements of symmetry group, in Hopf-algebraic framework the quantum Lie algebra describing quantum symmetry generators uniquely determines by duality the quantum symmetry group.

In this note we shall consider mostly the consequences of Hopf-algebraic $\kappa$-deformed Poincar\'{e} symmetries framework, introduced in 1991   \cite{lukchin3}. This algebraic framework was a conceptional basis of proposed in 2001 \cite{lukchin8,lukchin9} Doubly Special Relativity (DSR), with two invariant parameters $c$ and $\kappa$. The aim of DSR framework is to describe the modification of relativistic kinematics as well as to study the astrophysical effects which could reveal the need for modification of special relativity theory. In DSR considerations usually the studied are not restricted by the rigidity of Hopf-algebraic framework, in particular only the part of DSR authors assumed that space-time coordinates are not commutative (see e.g. \cite{lukchin10}).

Below, in Sect. 2., we shall list briefly the modifications of special relativity formulae. In Sect. 3 we shall comment on the deformation of field-theoretic framework, and relate via the formalism of star product the noncommutative and commutative fields. In Sect. 4 we shall list briefly the ideas which led to noncommutative gravity theory. It appears that the corrections to Einstein action is proportional to 
$\frac{1}{\kappa^2}$, i.e. of second order in the deformation parameter.

\section{Modification of special relativity}

The essence of Einstein`s introduction of special theory of relativity is the modification of Galilean symmetries into Poincar\'{e} symmetries, described by the transformations of the Poincar\'{e} group ($a_\mu, \Lambda_\mu^{\ \nu}$)
\begin{equation}\label{luchi3}
x'_\mu = \Lambda_{\mu}^{\ \nu} \, x_\nu + a_\mu
\qquad
\Lambda_{\mu}^{\ \nu} \,
\Lambda_{\nu}^{\ \rho}
= \delta_{\mu}^{\ \rho} \, .
\end{equation}
The symmetries 
(\ref{luchi3})  
are generated by the Poincar\'{e} algebra ($P_\mu , M_{\mu\nu}$) with mass and spin
Casimirs ($m \in [ 0, \infty)$, $s=0,\half, 1 \ldots$)
\begin{equation}\label{luchi4}
P_\mu P^\mu = - m^2 
\qquad 
W_\mu W^\mu = m^2 \, s(s+1) \, .
\end{equation}
Poincar\'{e} algebra is a Hopf algebra with Abelian 
(primitive) coproducts; from the last property follows e.g. that for relativistic systems we get Abelian addition law of fourmomenta
\begin{equation}\label{luchi5}
p^{(1+2)}_\mu = p^{(1)}_\mu + p^{ (2)}_\mu
\end{equation}
Relation (\ref{luchi5}) describes the fourmomentum of free two-particle system (1+2) composed as the tensor product of single particle states (1) and (2).

In $\kappa$-deformed special relativity the formulae (\ref{luchi3}-\ref{luchi5}) describing free relativistic particles are modified. If we replace $x_\mu$ by $\widehat{x}_\mu$ (see 
(\ref{luchi1}), (\ref{luchi2})) the relations (\ref{luchi3}) remain valid, but with the $c$-number Poincar\'{e} group parameters ($a_\mu , \Lambda_\mu^{\ \nu}$) replaced by operator-valued quantum Poincar\'{e} generators $(\widehat{a}_\mu , \widehat{\Lambda}_\mu^{\ \nu})$  \cite{lukchin11,lukchin4,lukchin12}.
The $\kappa$-deformation of Poincar\'{e} algebra depends on the choice of the basic generators; usually it is used the basis with classical Lorentz algebra sector \cite{lukchin4,lukchin13}, with modified only one Poincar\'{e} algebra commutator \cite{lukchin13,lukchin14}.
\begin{equation}\label{luchi6}
[N_i , P_j ] = i \delta_{ij} 
\left[\frac{\kappa}{2}(1 - e^{- \frac{2P_0}{\kappa}}
+ \frac{1}{2\kappa}\, \vec{P}^{\ 2}\right]
- \frac{i}{\kappa} \,P_i P_j \, .
\end{equation}
The modification (\ref{luchi6}) leads to the following $\kappa$-deformed mass Casimir
\begin{equation}\label{luchi7}
P_\mu P^\mu \to 2 \kappa \left(\sinh \frac{P_0}{2\kappa} \right)^2
- \vec{P}^{\ 2}
\end{equation}
and the nonAbelian three-momentum addition law:
\begin{equation}\label{luchi8}
P^{(1+2)}_i = P^{(1)}_i \,
e^{- \frac{P_0^{(2)}}{2\kappa}}
+
P^{(2)}_i \,
e^{- \frac{P_0^{(2)}}{2\kappa}}\, .
\end{equation}
Change (\ref{luchi7}) and (\ref{luchi8}) leads to important consequences e.g. 

1) classical relativistic energy momentum dispersion relation $E_{el} = (\vec{p}^{\ 2} + m^2)^{\half}$ is modified
\begin{equation}\label{luchi9}
E = \kappa \, c\ \hbox{arcsinh}
\frac{(\vec{p}^{\ 2} +m^2)}{\kappa c}
= E_{cl} + {\cal O} (\frac{1}{\kappa^2})
\end{equation}
and $E < E_{max}$ where $E= E_{max} = \kappa \, c$ in 
$|\vec{p}| \to \infty$ limit.

2) The notion of light-cone should be ceased, and the velocity of massless particles (e.g. photons) approaches $c$ only if $|\vec{p}|\to \infty$.

3) The Lorentz transformations in momentum space preserving the mass-shell conditions are modified
  \cite{lukchin15,lukchin16,lukchin10}.
 
4) Important astrophysical effects, in particular the appearance  of GKZ threshold and the evolution of the Universe in inflation period are changed {\cite{lukchin17,lukchin18,lukchin19}}.

In order to describe $\kappa$-deformed kinematics one has to extend further the $\kappa$-deformation to the phase space, what is provided by so-called Heisenberg double construction \cite{lukchin7,lukchin20}. In $\kappa$-deformed phase space the fourmomenta coordinates $p_\mu$ remain commutative, but the canonical PB implying standard Heisenberg algebra relation is changed. In such a way we obtain noncanonical symplectic structure which leads to the modification of classical and quantum Hamilton equations, 
what implies the modified version of QM.

\section{From noncommutative to commutative framework: the $\star$-product formalism  }

It is known from quantum mechanics that one can represent noncommutative quantum observables in terms of notions of commutative geometry (matrices, differential operators...). Analogously one can realize the non commutative fields on quantum  Minkowski space by classical commutative Minkowski fields
\begin{equation}\label{luchi10}
\varphi(\widehat{x}) 
{\xrightarrow[\hbox{Weyl map}]{\phantom{sssslhhhhhhhhllllss}}}
\varphi({x}) = \Omega(\varphi(\skew{-1}\widehat{x}))
\end{equation}
provided that we represent correctly the multiplication of noncommutative fields by so-called $\star$-product (see e.g. \cite{lukchin21})
\begin{equation}\label{luchi11}
\varphi(\widehat{x}) \chi(\skew{-1}\widehat{x})
{\xrightarrow[\phantom{x} ]{\phantom{sssslhhhhhhhh}}}
\varphi({x}) \star \chi({x}) \, .
\end{equation}
The map (\ref{luchi11}) is a homomorphism, i.e. preserves all algebraic relations satisfied by fields $\varphi(\widehat{x})$. Such map
is nonlocal and highly non-unique, due to the ordering prescription which should be added  in order to specify in complete way the field operator $\varphi(\widehat{x})$.

If the momentum sector of the phase space  remains after deformation  commutative, one can express the general Weyl map (\ref{luchi10}) in terms of the Weyl map of Fourier exponentials
\begin{equation}\label{luchi12}
\Omega(\varphi(\widehat{x})) =
\int d^4 p \, \widetilde{\varphi}(p) 
\Omega(e^{ip\widehat{x}})
\end{equation}
and the $\star$-product (\ref{luchi11}) is determined by the general formula
\begin{equation}\label{luchi13}
e^{ipx} \star\, e^{iqx} = \widehat{O}(x, \frac{\partial}{\partial x},
y, \frac{\partial}{\partial x}) e^{ipx} \, e^{iqy}\Big|_{x=y} \, .
\end{equation}
The nonlocal operator $\widehat{O}$ depends on the type of deformation. For simplest canonical deformation (formula (\ref{luchi1}) with $\theta_{\mu\nu}^{\quad \rho} = 0$) it is represented by a Moyal star product $\star_{0}$
\begin{equation}\label{luchi14}
\widehat{O}_{(0)} = \frac{i}{2} \, \frac{\partial}{\partial x_\mu}
\theta_{\mu\nu} \, \frac{\partial}{\partial y_\nu}
\end{equation}
and leads to quite simple formula
\begin{equation}\label{luchi15}
e^{ipx}  \star_{(0)} \, e^{iqx} =
e^{ipx} \, e^{iqx} \, e^{\frac{i}{2}\, p^\mu
\theta_{\mu\nu} \, q^\nu} \, .
\end{equation}
For the Lie-algebraic deformation (formula (\ref{luchi1}) with $\theta_{\mu\nu}=0$) due to BCH (Baker-Campbell-Hausdorff) operator formula
\begin{equation}\label{luchi16}
e^{i \alpha^\mu \, \widehat{x}_\mu} \,\cdot \,
e^{i \beta^\mu \, \widehat{x}_\mu} 
= e^{i \gamma^\mu(\alpha, \beta) \, \widehat{x}_\mu} 
\end{equation}
where $\gamma^\mu(\alpha, \beta)= \alpha^\mu + \beta^\mu
+ \theta_{\rho\tau}^{\quad \mu} \, \alpha^\rho \beta^\tau
+ O(\theta^2)$, we get
\begin{equation}\label{luchi17}
e^{ip^\mu x_\mu}
\star_{(1)} e^{iq^\mu x_\mu}
= e^{i \gamma^\mu(p,q) x_\mu}
\end{equation}
and
\begin{equation}\label{luchi18}
\widehat{O}_{(1)} = \exp i x_\mu \gamma^\mu (\frac{\partial}{\partial x},\frac{\partial}{\partial y})\, .
\end{equation}
The $\kappa$-algebra (\ref{luchi2}) is a soluble example of Lie-algebraic structure, and one obtains the closed  formula the function
$\gamma^\mu (p,q)$ \cite{lukchin22,lukchin23}
\begin{equation}\label{luchi19}
\gamma^{0} = p^{0} + q^{0}
\qquad
\gamma^i= \frac{
f_\kappa(p^0) e^{\frac{q_0}{\kappa}}
p^i + f_\kappa(q_0) q^i}{f_\kappa(p^0 + k^0)}
\end{equation}
where $f_\kappa(x) = \frac{\kappa}{x}(1-e^{\frac{x}{\kappa}})$.

Using $\star$-product formalism one can represent the noncommutative field theory as the standard local field theory with nonlocal $\star$-multiplication rule. In the simplest case of canonical deformation (see (\ref{luchi14})--(\ref{luchi15})) the modification of standard theory is milder, because the mass Casimir (see (\ref{luchi4})) and the theory of free fields is not deformed. The Abelian addition law for the fourmomenta remain valid, and the nonlocal structure in interacting theory is introduced by the phase factor $\exp \half p \theta q$
 (see (\ref{luchi15})) entering into the Feynmann diagram vertices  \cite{lukchin24}.   Every well-known model of classical field theory (e.g. gauge theories, QED, QCD, Einstein gravity) has been recently canonically deformed in the literature.
 
 The Lie-algebraic deformation leads to more complicated modifications. For mostly studied $\kappa$-deformation (\ref{luchi2}) the substitution (\ref{luchi7}) leads to the modification of free field equations and correspondingly modified propagators in $\kappa$-deformed perturbation theory (see e.g. \cite{lukchin25}). The Abelian conservation law (\ref{luchi5}) of four-momenta are changed; with suitable choice of fourmomentum variables one gets the $\kappa$-deformation addition law (\ref{luchi8}). At present the interacting $\kappa$-deformed field theory which is covariant under the $\kappa$-deformed relativistic symmetries is still under construction.
 
 \section{From noncommutative framework to modified gravity theories}
 
 In the literature mostly the canonical deformation of gravity has been proposed \cite{lukchin26,lukchin27}. After introducing $\star$-multiplication (\ref{luchi14}--\ref{luchi15}) in order to describe the reparametrization-invariant deformed gravity one should consistently modify the differential calculus as well as the structure of diffeomorphisms.
 
 One introduces
 \\
 i) $\star$-deformed differential calculus
 \\
 ii) $\star$-deformed composition rule of diffeomorphisms 
 $\delta_\xi = \xi^\mu \partial_\mu$
 \begin{equation}\label{luchi20}
 [ \delta_\xi , \delta_\eta ] = \delta_{\xi \star \eta}\, .
 \end{equation}
It should be observed that the algebraic commutator of diffeomorphisms is not deformed, but the standard Leibniz rule is changed into the modified one. Using the notion of coproduct of diffeomorphisms, the classical primitive coproduct is modified into the well-defined noncocommutative one.

In gravitation theory if our basic fields are vierbeins $e^a _\mu$, the metric field is deformed as follows:
\begin{equation}\label{luchi21}
g_{\mu\nu} = e^a _\mu \, e_{\nu a} \to \widetilde{g}_{\mu\nu}
= \half e^a_{(\mu} \star e_{\nu b)} \, .
\end{equation}
Introducing the determinant  $\det_{\star_{(0)}}(e^a_\mu)$ with standard multiplication of vierbeins replaced by $\star$-product the deformed Einstein action takes the form \cite{lukchin26}
\begin{equation}\label{luchi22}
S_{E} = \frac{1}{2\kappa^2}
\int d^4 x \, \det{}_\star (e^\mu_a) \star \widetilde{R}
\end{equation}
where using the $\star$-multiplication in the definition of scalar curvature in terms of basic vierbeins one can expand  the modified scalar curvature $\widetilde{R}$ in terms of deformation parameters $\theta_{\mu\nu}$ as follows
\begin{equation}\label{luchi23}
\widetilde{R} = R + \theta^{\alpha\beta} R^{(1)}_{\alpha\beta}
+ \half \theta^{\alpha\beta} \theta^{\gamma\delta}
R^{(2)}_{\alpha\beta\gamma\delta} + \ldots \, .
\end{equation}
An important result has been shown \cite{lukchin28,lukchin29}
that only second order corrections to the scalar curvature $R$ are nonvanishing, i.e.
\begin{equation}\label{luchi24}
R^{(1)}_{\alpha\beta} = 0 \, .
\end{equation}
Using other technical tool of Seiberg-Witten transformation it has been shown \cite{lukchin30} that the relation (\ref{luchi24}) is valid as well for $\kappa$-deformation.

The modified second order Einstein action (\ref{luchi22}) is nonlocal and invariant under deformed diffeomorphisms parametrized in D=4, as in standard gravity, by four functions $\chi_\mu$. 
It should be added that the modified Einstein action can be equivalently obtained if we use first order formulation \cite{lukchin31} and the geometric Mansouri-McDowell 
approach~\cite{lukchin32}. 
Other class of deformed gravity theories are obtained if we keep the standard Leibniz rule for diffeomorphisms. In such a case the theory is invariant only under the subclass of general coordinate transformations and describes the deformation of unimodular gravity~\cite{lukchin33}.

\subsection*{Acknowledgments} The author would like to thank the organizers of 1st Galileo-Xu Guangqi Meeting in Shanghai for their warm hospitality. The paper was supported by Polish Ministry of Science and Higher Education grant NN202318534.


\end{document}